\documentclass[final]{aipproc}

\layoutstyle{6x9}

\SetInternalRegister\hbadness{8000} 

\def\apj{ApJ}
\def\apjl{ApJ}
\def\aap{A\&A}
\def\mnras{MNRAS}
\def\nat{Nature}
\def\physrep{Phys. Rep.}

\newcommand\doingARLO[2][]{%
  \ifx\mmref\undefined #1\else #2\fi
}

\begin{document}

\title 
      [Progenitors of Long Gamma-ray Bursts]
      {Progenitors of Long Gamma-ray Bursts}


\author{Melvyn B. Davies}{
  address={Lund Observatory, Box 43, SE-221 00 Lund, Sweden},
  email={mbd@astro.lu.se},
}

\iftrue
\author{Andrew J. Levan}{
  address={Department of Physics, University of Warwick, Coventry, CV4 7AL},
  email={a.j.Levan@warwick.ac.uk},
  altaddress={Lund Observatory, Box 43, SE-221 00 Lund, Sweden}
}

\author{Josefin Larsson}{
  address={Institute of Astronomy, Madingley Road, Cambridge CB3 0HA, UK},
  email={jlarsson@ast.cam.ac.uk},
  altaddress={Lund Observatory, Box 43, SE-221 00 Lund, Sweden}
}

\author{Andrew R. King}{
  address={Department of Physics and Astronomy, University of Leicester, 
Leicester LE1 7RH, UK},
  email={ark@ast.le.ac.uk},
}

\author{Andrew S. Fruchter}{
  address={Space Telescope Science Institute, 3700 San Martin Dr., Baltimore,
MD 21218, USA},
  email={fruchter@stsci.edu},
}

\fi

\copyrightyear  {2007}

\begin{abstract}
Pinpointing the progenitors of long duration gamma-ray bursts (LGRBs)
remains an extremely important question, although it is now clear that
at least a fraction of LGRBs originate in the core collapse of
massive stars in type Ic supernovae, the pathways to the production
of these stars, and their initial masses, remain uncertain. Rotation
is thought to be vital in the creation of LGRBs, and it is likely
that black hole creation is also necessary. We suggest that 
these two constraints
can be met if the GRB progenitors are very massive
stars $>$ 20 M$_{\odot}$ and are formed in tight binary systems. 
Using simple
models we compare the predictions of this scenario with
observations and find that the location of GRBs on their
host galaxies are suggestive of main-sequence masses in 
excess of 20 M$_{\odot}$, while 50\% of the known 
compact binary systems may have been sufficiently 
close to have had the necessary rotation rates for GRB
creation. Thus, massive stars in compact binaries are a likely
channel for at least some fraction of LGRBs.
\end{abstract}

\date{\today}

\maketitle

\section{Introduction}

The link between some long-duration gamma-ray bursts (LGRBs) and stellar
collapse is now firmly established 
\cite{1993ApJ...405..273W, 2003Natur.423..847H, 2003ApJ...591L..17S}
In particular LGRBs appear to originate in type Ic
supernovae and frequently in hypernovae.  There is 
significant diversity in the LGRB population, including the highly
luminous (``classical") burst population, originating from
a mean redshift of 2.8 \cite{2006A&A...447..897J}, 
a nearby lower 
luminosity sub-class, which has a much higher space density
(e.g. \cite{2006Natur.442.1011P}), and possibly some LGRBs which are
not associated with supernovae.
This suggests there may be some variety in the progenitors of the bursts,
and that they may not come from systems undergoing identical evolution. Nonetheless,
there remains a strong drive to identify the progenitor stars themselves.

In the standard picture, core-collapse supernovae lead to LGRBs when
the stellar core has a rapid rotation just prior to collapse. 
Some of the infalling material then forms a torus around the
newly-formed central compact object (often assumed to be a black hole)
and subsequent accretion of the material in the torus then
fuels the LGRB.
We are naturally left with two questions: 1) What leads to the 
high rotation rates required to form a torus? 2) What is the 
minimum progenitor mass required (or, in other words is the formation
of a black hole (BH) rather than a neutron star (NS) necessary)?

For massive single stars, it is not clear
whether sufficiently high central rotation rates may be maintained to
produce a torus on core collapse (e.g. \cite{1999ApJ...524..262M, 2005A&A...435.1013P}), although 
it has been suggested that rapidly rotating metal--poor stars can
retain sufficient angular momentum 
\cite{2005A&A...443..643Y, 2006ApJ...637..914W}
Alternatively, binary scenarios suggest a way of removing the
envelope and providing a source of angular momentum (e.g. 
\cite{2004MNRAS.348.1215I, 2004ApJ...607L..17P}).
We have explored this idea
further \cite{2006MNRAS.372.1351L} and present a summary of the key results
here.

In addition, we have explored whether the observation of 
\citet{2006Natur.441..463F}
that LGRBs are more concentrated on their host galaxy light
than SNe may be due to the greater mass of a LGRB progenitor
than a typical SNe progenitor \cite{XXXLarssonetal}. A review
of the key results of this work is also presented here.

\section{Neutron star binaries and long-duration gamma-ray bursts}

\subsection{Evolutionary pathways to compact object binaries}

Binaries containing two compact objects (ie white dwarfs, neutron
stars or black holes) can be formed through several channels
(see e.g. \cite{2002ApJ...572..407B}). The basic scheme is shown in
Fig. 1, and is essentially the same as that described in
\citet{1991PhR...203....1B}.

The primary evolves first to produce either a black hole or neutron
star, possibly following a phase of mass transfer to the secondary.
When the secondary evolves, and fills its Roche lobe, the formation
of a common envelope phase is likely owing to the mass ratio of the
system. The black hole/neutron star and the helium-star core of the secondary
will then spiral towards each other as the enshrouding gaseos 
envelope is ejected. The post-common envelope system will be a compact
binary. A key point is that the helium-star core will be rotating rapidly
because of tidal locking. In at least some systems, the rotation rate
will be sufficiently large that when the secondary explodes as 
a core-collapse supernova, some of the infalling material will form
a torus around the newly-formed neutron star or black hole.

\begin{figure}
  {\includegraphics[height=0.75\textheight]{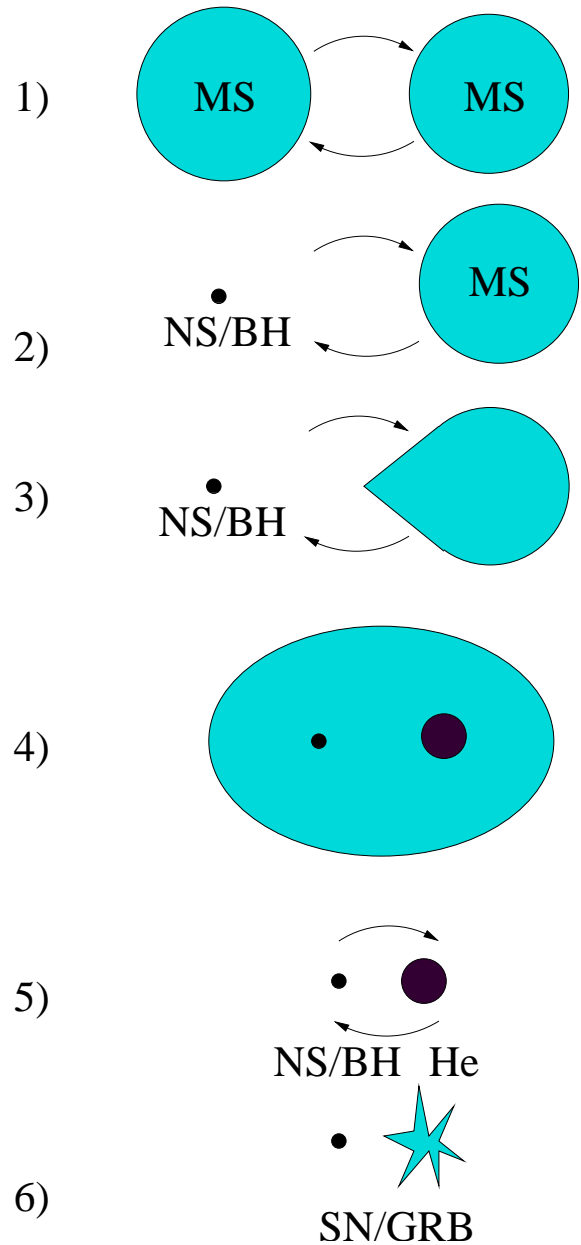}
  \vspace{0.5truecm}}
  \caption{An evolutionary pathway to the creation of a
binary containing a rapidly-rotating core-collapse supernova
in a tight orbit.
The primary evolves first, possibly transferring material to
the secondary (stage 1). It then produces a neutron star (NS)
or black hole (BH),
when it explodes as a core-collapse supernova (stage 2).
The secondary then evolves, filling its Roche lobe (stage 3) and
transferring material to the NS/BH producing a common envelope
phase (stage 4). The NS/BH and He core of the secondary spiral
together ejecting the surrounding envelope producing a very
compact binary (stage 5). Tidal locking produces a rapidly-rotating
He star such that the rotation is significant when the secondary
explodes as a core-collapse supernova, with a torus being formed
around the central compact object by infalling material.}
\end{figure}

\subsection{Discs around neutron stars and black holes}

In a core-collapse supernova, the material within the stellar
core collapses to form a black hole or neutron star.
If a core of a massive star is rotating sufficiently rapidly, 
some of the infalling material will be centrifugally supported
(as it will carry its angular momentum with it) and form a torus
around the central black hole or neutron star. The torus
will have a radius of $D G M_{\rm c} / c^2$ 
(where $M_{\rm c}$ is the mass of the black hole or neutron
star) if the specific angular momentum of the infalling material
 is given by $\sqrt{D} G M_{\rm c} / c$. In order to have a stable
orbit around a (non-rotating) black hole, we require $D \ge 6$
(or a radius of around 12 km for a 1.4 $M_{\odot}$ black hole).  
Forming a disc around a proto-neutron star requires more angular
momentum as a disc would have to be formed with a radius somewhat
larger than 10 km.

If we assume that tidal locking occurs at the
beginning of the helium main sequence and that for the late stages of
evolution of the helium star the core
decouples from the envelope.  In this case, following 
\citet{2004ApJ...607L..17P}
we can compare the required angular momentum at the time of
collapse with that at the edge of the iron core at the start of the
helium main sequence, and, assuming tidal locking equate this to an
orbital frequency (i.e.  $\omega = \sqrt{D} G M_{\rm c} / R_{\rm c}^2 c$, where
$R_{\rm c}$ is the radius of the iron core). Assuming synchronous rotation,
this gives a critical orbital separation \cite{2006MNRAS.372.1351L}

\begin{equation}
a < (4M_{\rm tot} c^2 R_{\rm c}^4 / 9DG M_{\rm c}^2)^{1/3},
\end{equation}
or
\begin{equation}
\left( {a \over R_{\odot}} \right) < {60 \over D^{1/3}} \left({R_{\rm c}
\over R_{\odot}}\right)^{4/3} \left({M_{\rm tot} \over
M_{\odot}}\right)^{1/3} \left({M_{\rm c} \over M_{\odot}}\right)^{-2/3},
\end{equation}
where $M_{\rm c}$ and $R_{\rm c}$ are the core mass and radius, and
$M_{\rm tot}$ is the total mass of the binary.  We use here
the models of \citet{2000ApJ...528..368H}, the key parameters are
$M_{\rm c} = 1.7$ M$_{\odot}$, $R_{\rm c}
\sim 0.1 R_{\odot}$, for a He star of mass 7.71 M$_{\odot}$, this
yields a critical separation of $\sim 3$ R$_{\odot}$ for the formation
of a disc at the innermost stable orbit of a 1.7 $M_{\odot}$ BH.

Do systems exist which are sufficiently tight to form discs
when the secondary undergoes a core-collapse supernova? 
In order to answer this question, one can consider the observed
compact binaries containing black holes or neutron stars
which would be formed after the second supernova explosion
(assuming the binary remained bound).
There are currently no known black hole-neutron star binaries.
However there are currently eight NS-NS binaries and three
WD-NS binaries. The semi-major axis and eccentricity of a binary
provides limits to the separation of the binary at the moment
of the second supernova. In Fig. 2 we plot the semi-major axes
and eccentricities of the observed systems. The plot also
illustrates the minimum separation required to form a disc
around the second compact object for discs of various radii.
From this figure, we can see that discs are likely to have formed during
the second supernova in about half of all osberved binaries.
Energy released from such discs may power some of the long gamma-ray
bursts, as discussed below.

\begin{figure}
  \includegraphics[height=0.5\textheight]{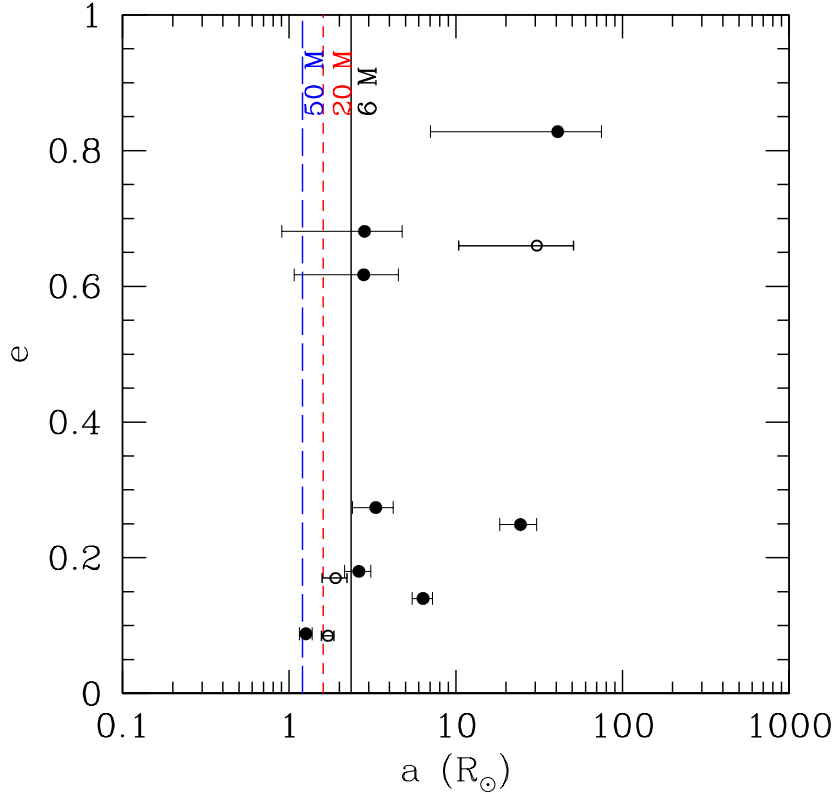}
  \caption{The eccentricity and semi-major axis of observed systems
containing either two neutron stars, or a neutron star and white
dwarf.  The error bars give an indication of the range of separations
between the two stars during their orbit. Neglecting the inspiral the
separation at the time of the supernova must be taken from this range
(see Table 1 for the calculations including the effects of
inspiral via gravitational radiation).
The three vertical lines represent the critical separations necessary
at the time of core collapse for a centrifugally supported disc to have
formed at a distance of 6M, 20M and 50M from the newly formed
compact object,  based on the models of
\citet{2000ApJ...528..368H}
described in section 3 (where $M= GM_{ns/bh}/c^2$ and is
$\sim 12, 40$ and 100 km for D=6,20  and 50 for a 1.4 M$_{\odot}$ NS). Lower
total masses, as were likely the case of J0737-3039 require
slightly tighter orbits, although the orbit is only a weak function
of the total mass ($M_{tot}^{1/3}$).
 If the binary separation is less than this (i.e.
to the left of the line) then disc formation is favoured.
The data are from \citet{2004MNRAS.350L..61C} and 
\citet{2005LRR.....8....7L}.
In cases where the masses of the two components
have not be measured 1.4 M$_{\odot}$ has been assumed. Known NS-NS
binaries are indicated with circles, while NS-WD or those with
uncertain companions are marked with open circles}.
\end{figure}

\subsection{The contribution of compact binaries to long gamma-ray bursts}

Long-duration gamma-ray bursts are commonly thought to originate
from discs around black holes located at the centre of a supernova.
However, one can also consider whether discs around newly-formed
neutron stars may also produce some form of gamma-ray burst.
The maximum energy released in the 
accretion  of the torus is given by $E = G M_{\rm ns/bh} M_{\rm acc} /
R_{\rm ns/bh}$,  or $E_{\rm ns} = 3.6 \times 10^{53} (M_{\rm acc} / M_{\odot})$
ergs, for a 1.4 M$_{\odot}$ NS and $E_{\rm bh} = 1 \times 10^{54} (M_{\rm acc}
/ M_{\odot})$ ergs, for any mass BH (since $M_{\rm bh} \propto R_{\rm bh}$).

The extrapolation from disc accretion to gamma-ray luminosity is far
from trivial since it requires an assumption about the conversion of
accretion luminosity into $\gamma$-ray energy. One plausible mechanism 
of providing this energy is via neutrino-antineutrino annihilation.
If this is assumed to be the energy source then the accretion energy  can
be related to the observed gamma-ray energy via several efficiency
factors which account for the conversion of accretion energy to
neutrinos, the cross section for neutrino - antineutrino annihilation,
the subsequent fraction of energy that is transferred into a baryon
free jet, and finally the fraction of this energy which is emitted as 
gamma-rays \cite{2006MNRAS.368.1489O}.  Following 
\citet{2006MNRAS.368.1489O}
we assume that the product of the these efficiencies is $\sim
10^{-3}$. Thus the observed luminosities of  low luminosity GRBs of
$10^{48} - 10^{50}$ ergs can be  explained by the accretion of $ 0.01
< (M_{acc} / M_{\odot}) < 0.3$ of material from the disc.

More-massive versions of the systems which formed the observed NS-NS
binaries are expected to form BH-BH binaries and
are thus candidates for classical long gamma-ray bursts.
Could both supernovae in such systems produce a gamma-ray burst?
One can consider the separation required in order to give the
core of the primary a sufficiently-high rotation rate. It is found that
the size of the stellar is restricted to less than a few solar 
radii. In other words, the primary of a binary containing two
massive stars is unlikely to be tidally spun-up sufficiently to produce
a gamma-ray burst. However the black-hole it produces will work
effectively to tidally spin-up the core of the secondary star
following a pathway similar to that shown in Fig. 1.

%

\section{A new constraint for gamma-ray burst progenitor mass}

\subsection{The basic idea}

 Recently \citet{2006Natur.441..463F}
have conducted a survey of the galactic
environments of both long duration GRBs and core collapse SNe (i.e.
all types of core collapse events, including SN II, Ib and Ic). These
results demonstrate that GRBs are highly concentrated on their host
light, significantly more so than the core collapse supernova
population.  \citet{2006Natur.441..463F}  further suggest that this can be
explained as being due to the GRBs originating in the most massive
stars, which, upon core collapse form black holes rather than neutron
stars. Here we further explore this possibility and attempt to derive
plausible limits on the progenitor lifetime and mass based on the
observed distributions of core collapse
 SNe and GRBs upon their host light. Using
a simple model, motivated by the distributions of young star clusters
in a local starburst galaxy, we explore the expected distributions of
stars of different masses upon their host galaxies and compare these
to the observed distributions from \citet{2006Natur.441..463F}. Our results
demonstrate that for plausible models more massive stars are {\it
  always} more concentrated on their host light than lower mass stars.
Further, given that supernovae originate from stars with initial
masses $>$ 8 M$_{\odot}$, we find that the observed distributions of
GRBs on their host galaxies can naturally be explained by progenitors
with initial masses in excess of 20 M$_{\odot}$.

\subsection{The model}

\begin{figure}
  \includegraphics[height=0.5\textheight]{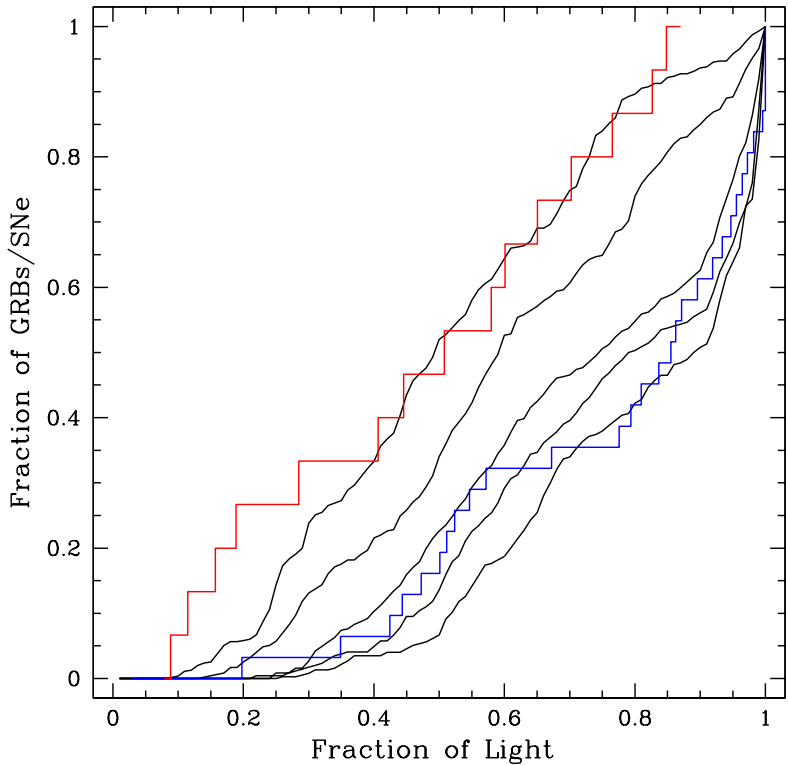}
  \caption{Fraction of objects plotted against fraction of light for
  observed SNe (red) and GRBs (blue) together with the results from
  our model (black lines). Black lines from top to bottom correspond
  to minimum progenitor masses of 8, 20, 40, 60, and 80
  $M_{\odot}$.}
\end{figure}

GRB host galaxies at high redshift are typically starburst
galaxies. The Antennae (NGC 4038/4039) are a natural, local,
analogue and were used as a template for constructing a simple
model for GRB host galaxies. In this model we consider
the population of young stellar clusters, as well as the 
stellar population contributing to the background light within
the galaxy.
The galaxy model consists of several key parameters:
the surface density of clusters; the distribution of cluster masses;
the distribution of cluster ages; the distribution of background 
light; and the distribution of clusters on the background light.
The first two properties were taken from observations of
NGC 4038/4039 when viewed at $z\sim1$.
The surface density of clusters expressed in terms of number of clusters
per pixel is $\sim 0.15$, although this is far from uniform across the
galaxy.  We use the observed distribution of cluster masses, which
follows $dN/dM_{\rm{cl}}\propto M_{\rm{cl}}^{-2}$, with
$M_{\rm{cl,min}}=4\cdot 10^4 \ M_{\odot}$ and $M_{\rm{cl,max}}=10^6\
M_{\odot}$, where $M_{\rm{cl}}$ is the cluster mass.  The age
distribution of clusters was taken from 
\citet{2005ApJ...631L.133F} and follows
$dN/d\tau \propto \tau^{-1}$, where $\tau$ is the cluster age. In
order to mimic an (almost) instantaneous burst of star formation,
clusters are created in the model according to this distribution over
a period of $10^7$ years. The distribution of the background light
adopted lies in the middle of the distributions observed for galaxies
hosting supernovae and gamma-ray bursts and is given by
$dN/dL_{\rm{pix}} \propto L_{\rm{pix}}^{-1.5}$ with
$L_{\rm{pix,max}}/L_{\rm{pix,min}}=20$, where $L_{\rm{pix}}$ is the
luminosity of a pixel. We distribute the stellar clusters
such that there is a correlation between background brightness and
the number of clusters present in a pixel. The degree of correlation
was chosen to give a good match to the cluster distribution observed
in NGC 4038/39.



\subsection{Results}

\begin{figure}
  \includegraphics[height=0.5\textheight]{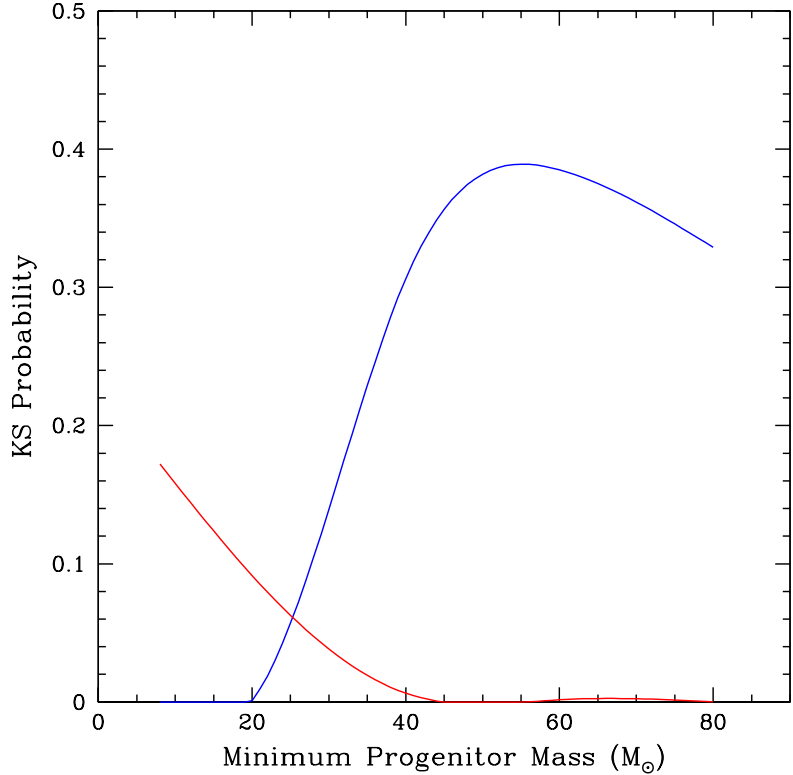}
  \caption{KS-probabilities of our model results following the observed
  SN (red) and GRB (blue) distributions. The probabilities are plotted
  as a function of the minimum progenitor mass and were calculated for
  masses of 8, 20, 40, 60, and 80 $M_{\odot}$. A spline has been
  fitted through the data points.}
\end{figure}

Using the parameters described in the previous section we performed
runs for minimum progenitor masses of 8, 20, 40, 60, and 80
$M_{\odot}$. The results are shown as black lines in Fig. 3
together with the observed distributions of SNe (in red) and GRBs (in
blue) from \citet{2006Natur.441..463F}.

The model distributions for all masses were KS-tested against the
observed SN and GRB distributions and the resulting probabilities are
shown as a function of mass in Fig. 4.  While the probability
of following the SN distribution decreases with increasing mass, the
likelihood of following the observed GRB distribution increases
rapidly from 8 to 40 $M_{\odot}$ and then flattens out, reaching a
weak maximum around 60 $M_{\odot}$. The shapes of the two probability
functions look the same for all realisations of the model, although
the peak probabilities can change by about 0.1 between different runs.
These results strongly suggest that GRB progenitors are significantly
more massive than SN progenitors.

\section{Summary}
We have suggested that GRBs originate from the most massive stars
\cite{XXXLarssonetal} and that many of these stars exist within tight
binary systems \cite{2006MNRAS.372.1351L}. After the GRB a NS-NS, NS-BH 
or BH-BH binary
may remain, as the separation at the time of the GRB is moderately small
many of the systems will merge within a Hubble time and might be observable 
as a short GRB. 
These systems are spectacular, at
some point in their evolution they will be observed as two supernova,
two GRBs as well as X-ray binary (and possibly even ultraluminous X-ray
source) phases.

We note that while it is likely that  LGRBs are formed via this channel 
it is not necessarily the only one operating, indeed the diversity in
the properties of observed GRBs implies there may be significant 
variations in the progenitors themselves, for example the low
luminosity burst population may be formed via neutron star creating
supernovae \cite{2006MNRAS.372.1351L,2006Natur.442.1014S}
while 
single stars, which undergo complete mixing on the main 
sequence (e.g. \cite{2006ApJ...637..914W}) 
may also create a subset of the GRBs. 

\begin{theacknowledgments}
MBD is
a Royal Swedish Academy Research Fellow supported by a grant from the
Knut and Alice Wallenberg Foundation. 
AJL is grateful to PPARC for a postdoctoral fellowship award. AJL also
thanks the Swedish Institute for support while visiting Lund.  
\end{theacknowledgments}



\end{document}